# High-Resolution, Three-Dimensional Reconstruction of the Outflow Tract Demonstrates Segmental Differences in Cleared Eyes

Susannah Waxman,*[1] Ralitsa T. Loewen,*[1] Yalong Dang,[1] Simon C. Watkins,[2] Alan M. Watson,[2] Nils A. Loewen[1]

*these authors have equally contributed

1: Department of Ophthalmology, University of Pittsburgh, Pittsburgh, Pennsylvania, United States of America
2: Center for Biologic Imaging and the Department of Cellular Biology, University of Pittsburgh, Pittsburgh, Pennsylvania, United States of America

**Word count:** 3501

**Grant information:** National Eye Institute K08EY022737 (NAL); Initiative to Cure Glaucoma of the Eye and Ear Foundation of Pittsburgh (NAL); Wiegand Fellowship of the Eye and Ear Foundation of Pittsburgh (YD); P30-EY08099 (NAL); Department grant by Research to Prevent Blindness (NAL).

# Abstract

**Purpose:** The rate of conventional aqueous humor outflow is the highest nasally. We hypothesized that this is reflected in regionally different outflow structures and analyzed the entire limbus by high-resolution, full-thickness ribbon-scanning confocal microscopy (RSCM).

**Methods:** We perfused pig eyes by anterior chamber cannulation with eight lectin-fluorophore conjugates, followed by optical clearance with benzyl alcohol benzyl benzoate (BABB). RSCM and an advanced analysis software (Imaris) were used to reconstruct a three-dimensional, whole-specimen rendering of the perilimbal outflow structures. We performed morphometric analyses of the outflow tract from the level of the trabecular meshwork (TM) to the scleral vascular plexus (SVP).

**Results:** Except for pigmented structures, BABB cleared the entire eye. Rhodamine-conjugated *Glycine max* agglutinin (soybean, SBA) labeled the outflow tract evenly and retained fluorescence for months. RSCM produced terabyte-sized files allowing for in silico dissection of outflow tract vessel at a high resolution and in 3D. Networks of interconnected lumina were traced from the TM to downstream drainage structures. The collector channel (CC) volumes were ten times smaller than the receiving SVP vessels, the largest of which were in the inferior limbus. Proximal CC diameters were up to four times the size of distal diameters and more elliptical at their proximal ends. The largest CCs were found in the superonasal and inferonasal quadrants where the highest outflow occurs.

**Conclusions:** RSCM of cleared eyes enabled high-resolution, volumetric analysis of the outflow tract. The proximal structures had greater diameters nasally while the SVP was larger in the inferior limbus.

**Introduction**

Intraocular pressure (IOP) is the only modifiable factor in glaucoma shown to slow progression of this leading cause of blindness.[1,2] In normal eyes, the trabecular meshwork (TM) was found to be a primary location of outflow resistance.[3–5] Sites distal to the TM have not been examined fully due to difficulties visualizing small structures within the sclera.[6,7] Recent TM bypass[8,9] and ablation studies[10–13] demonstrate that the outflow resistance downstream of the TM is much higher in eyes with glaucoma. Only a small fraction of patients (about 0.3%) achieve the predicted IOP equal to episcleral venous pressure.[13] Pre- and postoperative IOP is also correlated, suggesting an increased post-TM outflow resistance in patients with a higher preoperative IOP.[10] Unmasking the 3D architecture of these outflow structures is necessary to understand their anatomy and function. However, eyes that are more similar in size to humans' require image acquisition at a depth more than ten times greater compared to the commonly used mouse models with a scleral thickness around 40 micron[14–16] (human limbal sclera: 500 micron[17]). The technology and methods to detect small structures within solid tissues have only become available recently. Confocal microscopy is a powerful and versatile imaging modality[18,19] providing detection of a range of fluorophores at a high resolution and in the same sample.[20,21] Scatter and an absorbance of the exciting and emitted light within a tissue reduces the maximum depth of confocal microscopy to approximately 100 microns.[22] Scatter and absorption can be mitigated through tissue clearing approaches which remove light scattering and absorbing molecules and match the refractive index (RI) of the tissue to the mounting medium.[23–25] Clearing can increase the maximum depth of confocal imaging to many millimeters, often limited only by the microscope optics.[24] We recently showed in mouse brains that the depth and large area limits of traditional confocal microscopy can be overcome by employing a threefold strategy[24] that uses 1) a high numerical aperture long-working distance objective, 2) a tissue clearing technique,[23,24] and 3) ribbon-scanning to reconstruct large, two-dimensional images from which three-dimensional projections can be rendered.[24]

There are important differences between the human and the porcine outflow tract, including the lack of a Schlemm's canal.[26] However, the consistent high tissue quality, absence of naturally occurring glaucoma, short time from enucleation to usage, and wealth of functional outflow studies convinced us to leverage our pig eye model.[27–32] We applied ribbon-scanning confocal microscopy (RSCM) to image the lectin-labeled outflow tract anatomy throughout the limbal sclera. This approach poses high demands on data acquisition and processing but provides a comprehensive insight into the complex network of small structures that escape recently used spectral-domain optical coherence tomography[33,34] and outflow casting.[35] We hypothesized that this would allow correlating regional morphological aspects to the well-established outflow patterns in this species.[27–30,32] Lectins are ubiquitous, carbohydrate-binding proteins that have a high specificity for sugar moieties[36,37] and have been used to examine the glycosylation of the TM in glaucoma[38] but can also be employed to study vessels by binding to their glycocalyx.[39,40] To overcome the highly disordered alignment of extracellular matrix components in the sclera,[41] primarily water, collagen, and elastin, and their variable RIs,[41,42] we modified a benzyl alcohol benzyl benzoate (BABB) protocol as a clearing technique of the lectin-labeled anterior segment.[25,43] Using RSCM,[24] we reconstructed large volumes of the anterior segment and performed volumetric analysis of the outflow tract architecture.

## Materials and methods

### *Whole eye lectin perfusion*

Twenty-three freshly enucleated porcine eyes were obtained from a local abattoir (Thoma Meat Market, Saxonburg, PA, United States) and processed within two hours. After removal of extraocular tissue, the eyes were positioned facing up with the optic nerve stump secured in low-compression mounts (CryoElite Cryogenic Vial #W985100, Wheaton Science Products, Millville, NJ, United States). A perfusion with the pressure set to 15 mmHg by gravity was performed as previously described.[44] Briefly, the anterior chamber was cannulated with a 20-gauge needle positioned temporally and just anterior to the limbus. With the bevel facing up, approximately 100 µl of aqueous humor per eye was drained. Each lectin **(Table 1)** was explored twice in separate eyes. An intracameral bolus of 0.2 mg/ml of lectin in phosphate buffered saline (PBS) was injected followed by an infusion of the same lectin at a concentration of 0.02 mg/ml for 90 minutes. The lectins tested consisted of a Texas red-conjugated lectin, *Lycopersicon esculentum* agglutinin (TL; Texas red-conjugated; #TL-1176, Vector Laboratories, Inc., Burlingame, CA) and a rhodamine-conjugated lectin kit which consisted of Concanavalin A (ConA), *Dolichos biflorus* agglutinin (DBA), *Arachis hypogaea* agglutinin (PNA), *Ricinus communis* agglutinin (RCA), *Glycine max* agglutinin (SBA), *Ulex europaeus* agglutinin (UEA), and *Triticum vulgaris* agglutinin (WGA); (Rhodamine Lectin Kit I, #RLK-2200, Vector Laboratories, Inc., Burlingame, CA, United States). In addition to the lectin-perfused eyes, a control eye was perfused with PBS. An additional control eye was used to adjust and calibrate the infusion method to a physiological pressure with a pressure transducer (DTX-plus, Argon Medical Devices, Plano, TX; amplifier, ADInstruments, Colorado Springs, CO, USA).

The surface of the eyes kept was moist with PBS at all times. The eyes were perfusion-fixed with fresh 4% paraformaldehyde, (Catalog No. P6148, Sigma-Aldrich) post-fixed overnight, and hemisected equatorially into an anterior and a posterior segment. The sclera and cornea were left intact while the iris, ciliary body, lens, and vitreous were carefully removed as previously described.[31] Anterior segments were bisected into inferior and superior halves. The samples were protected from light to prevent photobleaching of fluorophores.

*Tissue clearing*

Five of the 23 eyes served to develop the tissue clearing technique. We focused on BABB as a clearing technique when pilot experiments with CUBIC (clear, unobstructed brain imaging cocktails and computational analysis)[23,24] caused the sclera to assume a gel-like consistency, expand, and gradually disintegrate. iDISCO (immunolabeling-enabled three-dimensional imaging of solvent-cleared organs)[45] was also evaluated but did not impart sufficient transparency. Lectin-perfused samples were transferred into borosilicate glass vials (Catalog No. 03-339-21E, Fisher Scientific, Hampton, NH), washed, dehydrated in increasing ethanol concentrations of 50%, 70%, 80% and 96% in PBS for 90 min each, and left in 100% ethanol overnight. The 100% ethanol was replaced daily for five days. A 1:2 mixture of benzyl alcohol (Catalog No. 305197, Sigma-Aldrich) and benzyl benzoate (Catalog No. B6630, Sigma-Aldrich) was created.[46] The dehydrated samples were transferred to a solution of BABB diluted 1:1 in ethanol, incubated for 24 hours, and transferred to 100% BABB for 30 minutes or until visibly clear **(Figure 1)**. For all incubations, samples were placed in solutions of at least twice the volume of the tissue and affixed to a vertical stage rotated at 20 revolutions per minute.

*Confocal microscopy*

**Screening with upright confocal microscopy**

We screened the signal intensity and regional binding specificity of the different lectins **(Table 1)** in cleared samples on a Fluoview FV1200 upright confocal microscope (BX61, Olympus, Tokyo, Japan). Mounting chambers were cut from vinyl sheet material (Grip Taupe Shelf/Drawer Liner, vinyl material, Model # 04F-C6U59-06, Home Depot) and affixed to glass slides with cyanoacrylate adhesive which formed a barrier to contain the BABB. The mounts were left to cure for a minimum of six hours to ensure the vinyl enclosure was firmly attached to the glass slide. After sample segments were positioned in their mounts, wells were filled with BABB and sealed on top with a coverslip; caution was taken to avoid trapping of air bubbles.

**Whole-specimen ribbon-scanning confocal microscopy**

After selecting SBA as the preferred lectin for our studies, we acquired complete limbal circumference scans of one eye. Following several pilots, one eye was comprehensively assessed. RSCM exceeded several days for a single specimen followed by the assembly of the volume and the analysis time which also exceeded several days. The scans were obtained on an RS-G4 ribbon-scanning confocal microscope (Caliber I.D., Andover, MA, United States) fitted with a Märzhäuser scanning stage (SCANplus IM 120 x 80 (#00-24-579-0000), Märzhäuser Wetzlar GmbH & Co. KG, Wetzlar, Germany). An Olympus 25X, 1.05NA water immersion objective (XLPLN25XWMP2, Olympus, Tokyo, Japan) was used to acquire volumes with a voxel size of 0.365 x 0.365 x 2.43 µm as previously described.[24] Images were acquired over a single channel with an excitation wavelength of 561 and an emission filter of 630/60. Laser percentage, high voltage (HV), and offset were held constant throughout the volume at 2, 72, and 19, respectively. Custom-designed mounting chambers consisted of two threaded anodized aluminum rings, two glass coverslips, and a silicone rubber gasket fitted to the thickness of each specimen. The bottom coverslip (40 mm, round) was sealed with vacuum grease, the chamber was filled with BABB, an appropriately sized gasket was inserted, a second cover glass was overlaid, and the top ring was screwed into place. The pressure between the coverslips secured the sample in place, and pressure on the gasket sealed the BABB solution inside the chamber. Due to the prolonged acquisition time, the ring mounts were secured in place in a custom-designed acrylic pool that allowed for a large volume of water to be placed under the objective. This enabled more than 24 hours of acquisition before requiring the addition of water due to evaporation.

### *Data processing and image analysis*

Volumes of the limbus scans were reconstructed with Imaris 9.0 software (BitPlane AG, Zurich, Switzerland) from full resolution, large area composite TIFF images generated by the microscope. All data were saved in the Imaris file format (.ims) using a Lempel-Ziv-Welch (LZW) compression algorithm resulting in files of 1.1 and 0.78 terabytes (TB). The data was stored on network file servers and accessed using 10 gigabits/s network infrastructure. The analysis was performed on custom-built workstations running an Intel Core i7 6700K processor, 64GB RAM, 10 gigabits/s networking and NVIDIA GeForce GTX 1070 graphics card. In each of the superior and inferior portions of the scan, ten contiguous, full-thickness representative surfaces of the limbal region were segmented into TM, collector channels (CC), and scleral vascular plexus (SVP) **(Figure 2 A3** and **B3)**. The TM was identified as a densely stained region of fibrous tissue at the base of each segment. CCs spanned from their site of connection to the TM to the SVP, a network of superficial vasculature that ran largely parallel to the episclera. Volumes of labeled outflow tract structures were derived from the Imaris statistics function **(Figure 3)**. The CC openings were counted and measured in the XY plane along the widest and narrowest lengths drawn through their central axes. The orifices were defined by the appearance of a bright wall in contrast to a dark lumen with defined edges, round or oval, and at least one traceable connection to a larger, more superficial, perpendicularly oriented SVP channel. All measurements were taken at the outermost point where parameters were met. Any orifices concealed by pigment, occasionally encountered in the pig, could not be visualized or considered. Quantitative data obtained from scans of the same eye were compared using the unpaired Student's t-test. A p-value lower than 0.05 was considered statistically significant.

**Results**

Within 20 minutes to an hour of the final BABB exposure, limbal tissue became transparent and allowed grid lines to be seen from behind a cleared anterior segment **(Figure 1)**. During screening with the laser-scanning confocal microscope, fluorescence was absent in cleared control eyes not perfused with lectins. SBA stained the TM intensely except for a few regions with a weak signal that corresponded to pigment. The eyes could be left for up to two months in BABB without a notable difference in transparency or fluorescent signal. The lectin-labeled, cleared anterior segments could be imaged via RSCM **(Figure 2)**. Scan acquisition times for the superior and the inferior limbus segments were 53.4 and 35.2 hours, respectively, while the corresponding cluster processor time per specimen was 23.8 and 12.2 hours. Data processing, segmentation, and analysis with Imaris took approximately 344 hours. The TL binding pattern matched that observed in lectin-stained cryosections made in previous studies.[47] Out of all screened lectins, only WGA and PNA discriminated between different outflow structures in RSCM **(Table 1)**. TL, RCA, and SBA labeled the outflow tract with the most uniform signal. We selected SBA as the primary lectin for this study due to these results and its availability.

RSCM showed irregularly shaped lumina embedded within the TM. Tributaries of a low signal connected and diverged within this tissue, running mostly parallel to the TM itself until they culminated in one or more ovoid lumina at the TM inner wall **(Figures 2 and Figure 3)**. The hinge-like orientation of these lumina resembled human outflow tract structures seen in electron microscopy cross-sections **(Figure 3)**.[48] Often, a single luminal structure was seen just distal to the convergence of three adjacent TM inner wall orifices. The superior hemisection fly-through demonstrated eight superotemporal and eleven superonasal CCs at the TM level that branched into sixteen superotemporal and twenty-five superonasal CCs at the SVP level **([Movie 1](Movie 1))**. The hemisection fly-through of the inferior limbus showed ten inferotemporal and twelve inferonasal CCs at the TM level and sixteen inferotemporal and twenty inferotemporal CCs at the SVP level **(Movie 2)**.

Moving from the proximal to the distal portion of the limbus, orifices at the TM inner wall gave rise to channels, analogous in structure to CCs and aqueous veins, that spanned from the TM up to the SVP with highly variable degrees of branching and tortuosity. All channels running vertically from the TM to the SVP were deemed CCs for this study. Lumina of some CCs could be traced to a single point of connection with the SVP just distal to the channel's origin at the TM. Other tracts were highly branched and connected to the SVP much farther from their opening at the TM level or in multiple places **(Figures 2 and Figure 3)**. Some filamentous structures in this region were too fine to distinguish a lumen even with an XY-resolution of 365 nm (**Movie 1** and **Movie 2**; **Supplementary Data 1**). CCs were connected to a single, branched, and circumferentially spanning outflow channel with luminal widths notably larger than all other drainage structures. This structure was located within the outer third of the sclera, distal to the TM, and ran parallel to the episclera; it contained many valve-like hinges that had a high fluorescent signal. Using Imaris as a visualization and analysis software for microscopy, automated surface reconstruction of the outflow tract was performed and surfaces were segmented, in silico, into TM, SVP, and CC regions **(Movie 3)**. Reconstruction of individual CC units, from their connection site at the TM to their connection site at the SVP, could be performed through manual tracing **(Movie 4)**. Relative volumes of outflow structures consistently demonstrate densely stained TM, a lower signal density in the region between the TM and SVP, and high signal density in distal, superficial tracts (**Figure 3**).

The CC volumes were more than ten times smaller than the SVP volumes (p<0.001, **Figure 4 A**). There were no differences between the temporal and nasal limbus, but the inferior CC average volume was 1.9 times the superior (p=0.002, **Figure 4B**) while the inferior SVP average volume was 1.4 times the superior (p<0.001, **Figure 4B**). The average cross-section area (**Figure 5A**) of the proximal end of CCs was 1.7 times larger nasally than temporally (p=0.035) and 4.1 times as large as the distal portion of the CCs (p<0.001). The largest proximal cross-sectional areas were seen in the superonasal and inferonasal quadrants (**Figure 5 B**). The cross-sections of proximal nasal and temporal CCs were 1.6-fold more elliptical (ratio of width and height) than their distal ends (p<0.001, **Figure 5C**). Collectors with the most elliptical shape at their proximal end were again found in the superonasal and inferonasal quadrant (**Figure 5D**).

**Discussion**

Outflow resistance in healthy eyes has traditionally been ascribed to the TM,[3,4] but recent laboratory experiments show about half to be located further downstream.[5,49] Our clinical studies of ab interno trabeculectomy (AIT), a plasma-mediated TM ablation, indicate that the outflow resistance downstream of the TM is much higher in eyes with glaucoma.[10–13] As four of our recent clinical AIT studies with up to 1340 patients[10–13] demonstrate, almost nobody achieves the episcleral venous pressure one would predict[13,50,51] and that pressures lower than 18.6 mmHg cannot easily be obtained without medications.[52] The cause and location of this distal outflow resistance are unknown. Several groups have hypothesized that valve-like structures at the collector orifices are the culprit,[49] but these are mostly disrupted in AIT suggesting an unidentified downstream location. To address these questions, we established a method to visualize and acquire large, full-thickness sections of the outflow tract including a volumetric reconstruction of the porcine outflow.

Various chemical clearing techniques have been developed to address similar questions of three-dimensional structure, connectivity, and the resulting organ function. Clearing methods can be grouped into four categories: solvent based (BABB; iDISCO), hyperhydration (Sca/e A2, CUBIC), simple immersion (FocusClear), and hydrogel embedding (CLARITY, PACT, PARS).[53] We chose pig eyes for this study for the abundance of outflow function data[27–32] and the short time from enucleation to perfusion, helping confine dyes to the intact vascular spaces. We found that a modified CUBIC protocol[24] did not impart the same level of scleral transparency as BABB, a reagent known to be effective in clearing tissues with a large content of extracellular matrix-like skin and gingiva.[54] We selected a BABB based protocol[55] in this study because it empirically yielded the most transparent samples. Also, BABB is relatively easy and quick (days compared to weeks[54]), low cost, and preserved the signal from the lectin-conjugated fluorophore. Lectins have a long history of being used to label the glycocalyx of vascular endothelia.[39,56,57] We selected rhodamine-conjugated SBA for its uniform signal throughout the outflow tract. Compared to fluorophores with more blue-shifted excitation and emission maxima, the tissue has reduced autofluorescence and light absorption at the excitation and emission maxima of rhodamine.[45] These features lend themselves well to volumetric, full-thickness acquisition of the outflow tract. The speed and high resolution achieved with confocal ribbon-scanning enabled the characterization of large volumes of the outflow tract that would not have been feasible with traditional confocal or light-sheet imaging modalities.[21] This allowed us to readily isolate CCs and vessels of the SVP from within the complex outflow tract network. We found that the nasal CCs had the largest diameters and could be primarily found in the superonasal and inferonasal quadrant in between recti muscles **(Figure 5)**. In addition to this, the proximal openings are more elliptical than the distal ones in these quadrants, possibly indicating an expandable reserve capacity. This matches function studies well,[27,44] that showed that nasal outflow is higher than temporal outflow and that canalogram dye entry occurs preferentially in the superonasal and inferonasal anterior chamber angle.[27] In those experiments, time-lapse images of eyes before trabecular ablation had a more intense filling of the inferior SVP, matching the considerably larger volume in the present study **(Figure 4B)**. Pig eyes have an outflow tract similar to human eyes in several aspects (size,[26] giant vacuoles,[26] biochemical markers[58], increased nasal outflow[44]) but have Schlemm's canal-like segments instead of the mostly continuous single lumen of human eyes and a thicker TM.[26,59] Regardless, a similar, preferentially nasal and inferior flow can also be observed in human eyes.[60,61]

The volumes computed here demonstrate that CCs only contribute to a small spatial fraction of the complex outflow tract in porcine eyes and could generate a considerable post-trabecular outflow resistance. This resistance may occur close to the region of the CC-TM interface, where vessel diameters are large but collapsible and oval or closer to the SVP junction point where they are four times smaller **(Figure 5A)**. Serial sections at the level of the TM and CCs showed numerous fine, filamentous structures that appeared to be connected on only one side and might indicate lymphatic sacs or vascular sprouts.[62,63] Recent insights into outflow tract development have pointed towards a mixed origin of Schlemm's canal and downstream vasculature from lymphatic and blood vessels.[64–66] Although lectins show selective binding to specific sugar moieties, lectins that allow to reliably differentiate between lymphatic and blood vessels have not yet been identified. However, the differential expression of glycosaminoglycans,[67] hypotheses about flow properties, vessel wall adhesion, and mechanosensation[68–70] in glaucoma[38] can already be investigated using the lectins and techniques described here. Our whole-specimen approach of data acquisition and processing to understand the conventional outflow tract is demanding and poses challenges similar to hypothesis-driven omics integration.[71,72] The strategy presented here can now be applied to human eyes to identify the cause of post-trabecular outflow resistance in glaucoma.

In summary, we used pig eyes to establish a method that allows for investigation of structures in the distal outflow tract. A high-resolution, whole-specimen in silico rendering of the outflow tract could be achieved by combining BABB clearing, ex vivo administration of fluorescent lectins, and ribbon-scanning confocal microscopy. We found that CCs had more elliptical openings, larger proximal diameters, and increased volumes in the superonasal and inferonasal quadrant, while the SVP was largest inferiorly. This matches the areas of high outflow in prior function studies.

# Figures/Tables

*Table 1*

**Table 1.** Abbreviations, source of lectins, carbohydrate specificities, staining intensity and location (3 = strong; 2 = medium; 1 = weak; 0 = negative).

| | | | location | | |
|---|---|---|---|---|---|
| **lectin** | **source** | **monosaccharide binding** | TM | CC | SVP |
| tomato lectin TL | *Lycopersicon esculentum* tomato, fruit | β (1,4)-linked N-acetyl-glucosamine | 3 | 2 | 3 |
| concanavalin A ConA | *Canavalia ensiformis* Jack bean, seeds | α-man, α-glc | 2 | 2 | 2 |
| *Dolichos biflorus* agglutinin DBA | *Dolichos biflorus* horse gram, seeds | α-galNAc (n-acetylgalactosamine) | 3 | 3 | 3 |
| peanut agglutinin PNA | *Arachis hypogaea* peanut, seeds | gal β 3GalNAc or gal | 0 | 0 | 2 |
| *Ricinus communis* agglutinin RCA | *Ricinus communis* castor bean, seeds | d-galactose (Gal) | 3 | 3 | 3 |
| soybean agglutinin SBA | *Glycine max* soybean, seeds | α > β -galNAc (n-acetylgalactosamine) | 3 | 3 | 3 |
| *Ulex europaeus* agglutinin UEA | *Ulex europaeus* furze gorse, seeds | α-fucose (L-fucose) | 3 | 1 | 2 |
| wheat germ agglutinin WGA | *Triticum vulgaris* wheat, germ | glcNAc = f-acetylglucosamine or GlcNAc | 2 | 0 | 0 |

*Figure 1*

## perfusion-fixed anterior segment

before clearing    after clearing    flat mount of cleared eye

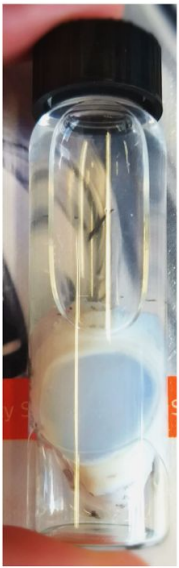
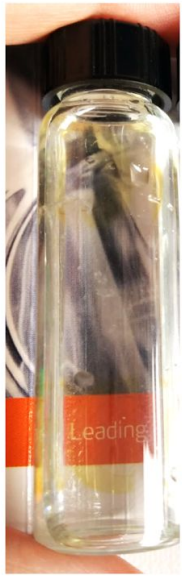
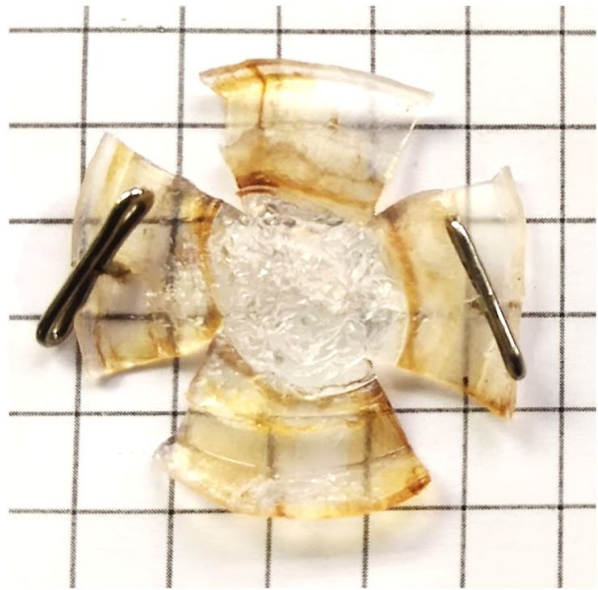

**Figure 1.** Macroscopic view of a BABB-cleared eye. Anterior segments of perfusion fixed eyes were cleared with BABB. Left: Anterior segment before and after clearing protocol. Right: An eye flat-mounted for demonstration purposes shows a high transparency except for parts that are heavily pigmented.

*Figure 2*

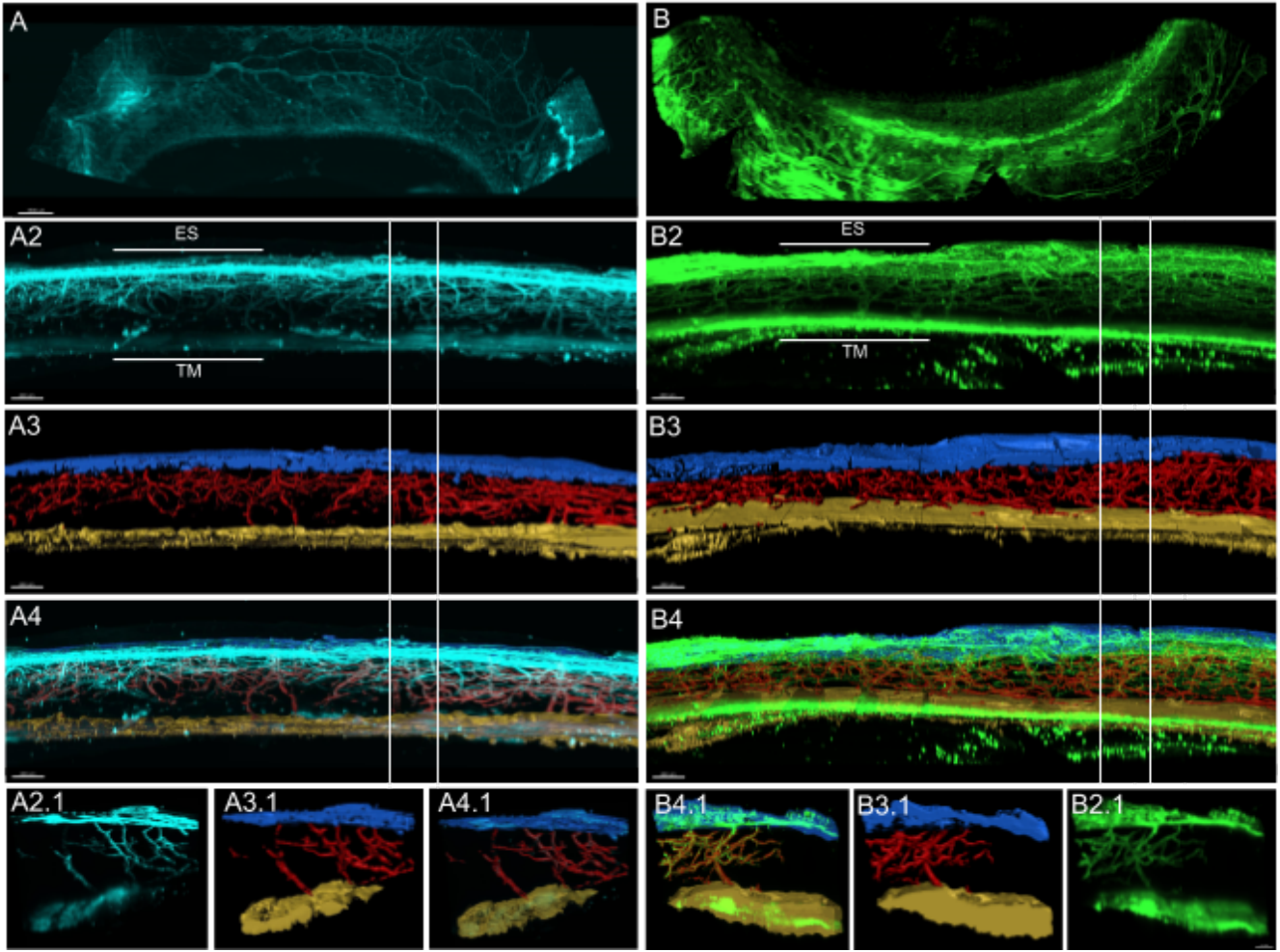

**Figure 2.** Volumetric limbal reconstruction. Soybean agglutinin (SBA)-labeled samples were visualized as maximum intensity projections (MIPs) of the superior (A, cyan) and inferior (B, green) limbus. MIPs A2 and B2 show fine elements of the aqueous humor outflow tract distal to the TM as seen from the anterior chamber angle. A3 and B3 show surface reconstruction and labeling of morphology with TM in yellow, CC in red, and SVP in blue. A4 and B4 show MIPs merged with corresponding surfaces. Vertical white lines demarcate boundaries within which representative sagittal subregions (A-B 2.1, 3.1, 4.1) are segmented. (TM=trabecular meshwork, ES=episclera)

*Figure 3*

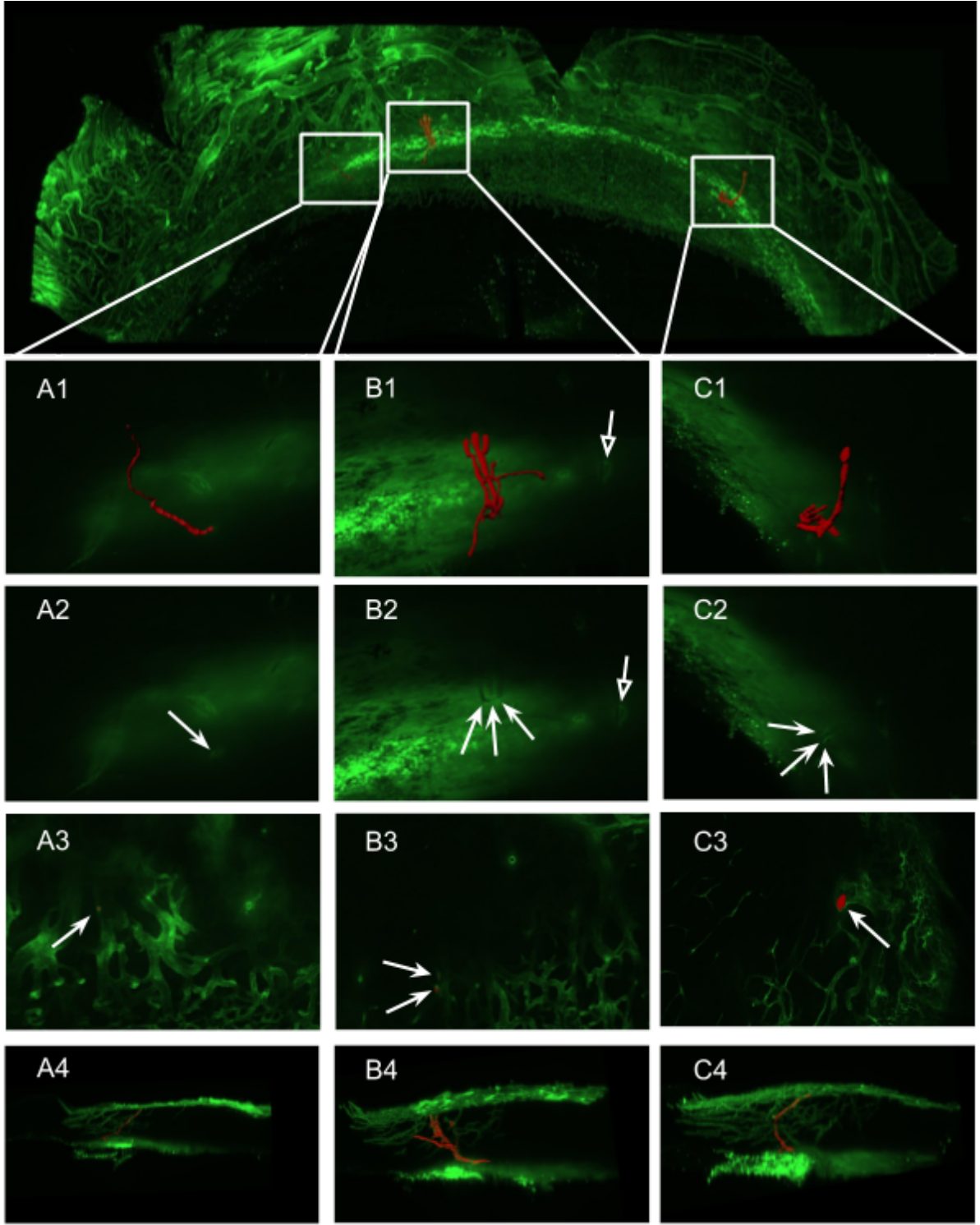

**Figure 3.** Collector morphology and connectivity from the TM to the SVP. The top panel shows an MIP of the inferior scan (green) with 3D reconstructions of representative collector channel lumina (red). Panels A1-C1 show the detail of delineated areas of the top panel at the depth of each channel's connection to the TM. A2-C2 show this site of connection (solid arrows). A3-C3 point to the site of the lumen's connection to the SVP (solid arrows) as vasculature begins to run parallel to the surface of the episclera. A4-C4 show sagittal sections of regions of interest, with an uninterrupted connection between TM and SVP. Hollow arrows mark hinge-like flaps by collector channel openings.

Figure 4

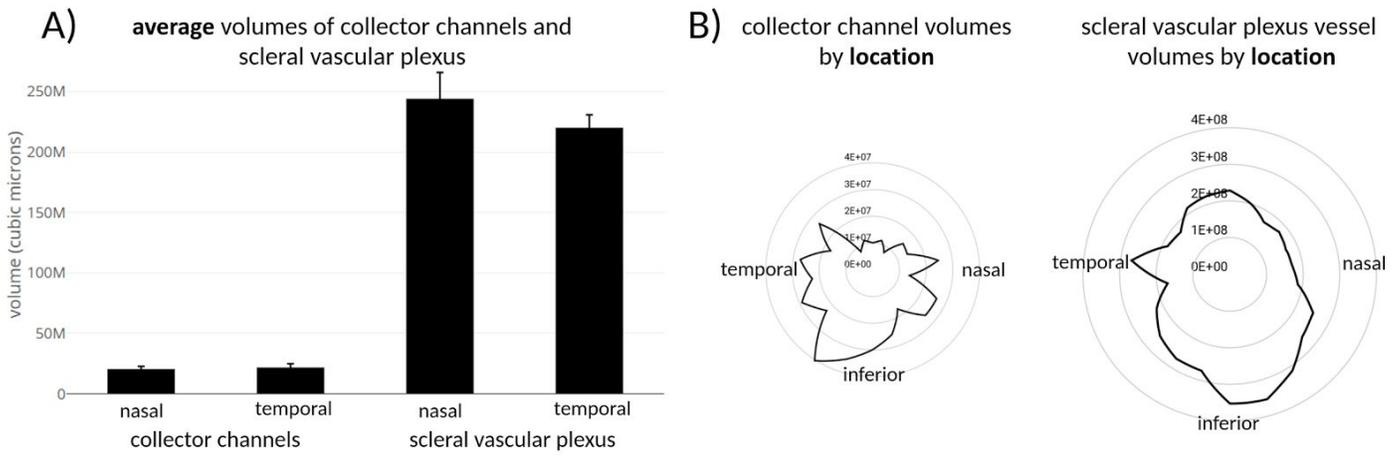

**Figure 4.** Volumes of collector channels (CC) and scleral vascular plexus vessels (SVP). A) Average volumes of SVP were about ten times larger than those of CC. B) Frontal view of a right eye. Larger CC and SVP volumes were found mainly in the inferior limbus.

**Figure 5**

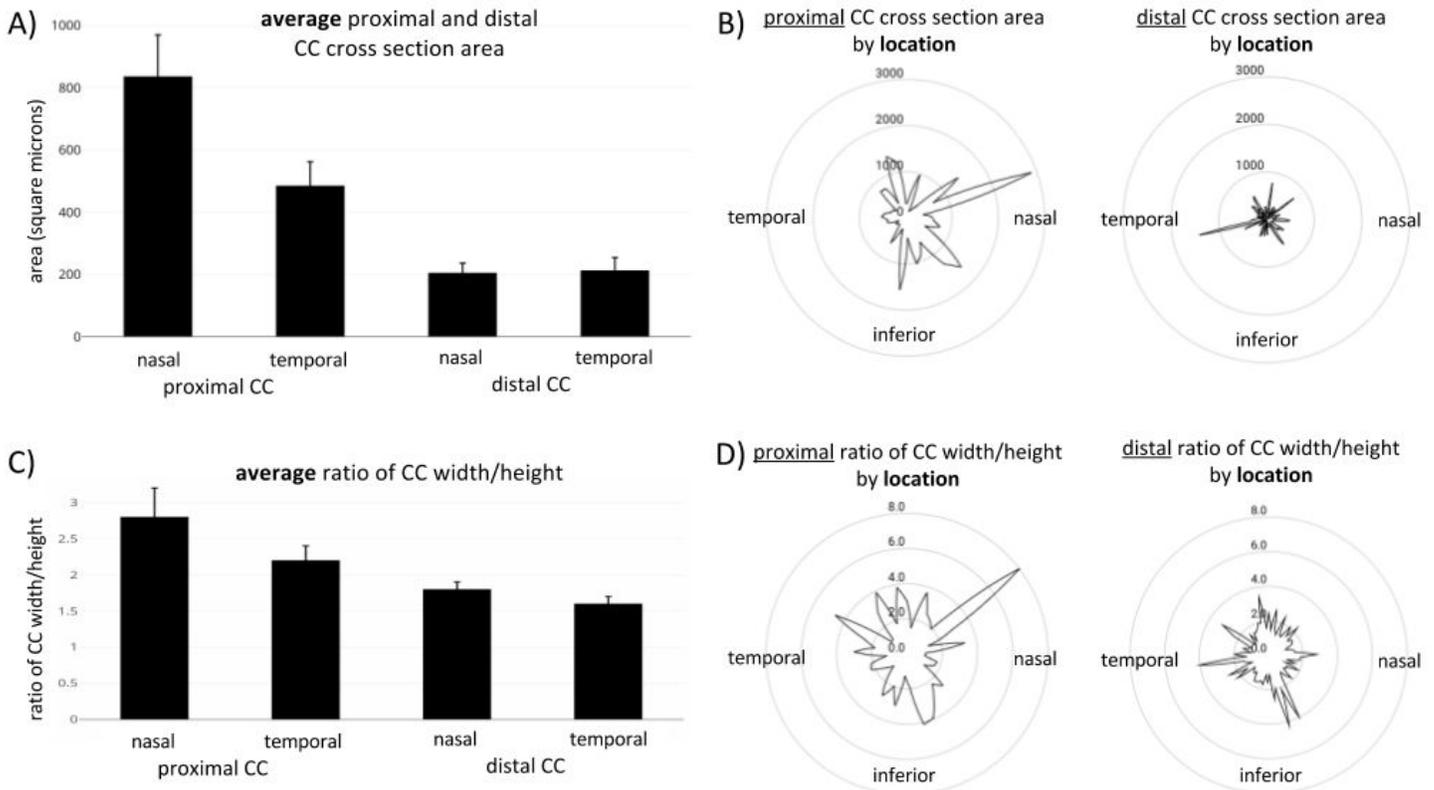

**Figure 5:** Cross-section area of collector channel sections. A) Collector channel openings at the level of the trabecular meshwork (proximal CC) had a 1.7-fold larger section area nasally than temporally (p=0.035). Both nasally and temporally, the proximal portion of CCs was larger than their distal portion. B) Frontal view of a right eye. The greatest section areas were found in the superonasal and inferonasal quadrant. C) Proximal nasal and temporal CCs were more oval compared to their distal ends as expressed by the ratio of the maximal to the minimal cross-section width. D) In the frontal view of this right eye, cross-section areas were more elliptical in the superonasal and inferonasal quadrant.

**Movies**

**Movie 1.** [Superior limbus fly-through.](#)

**Movie 2.** [Inferior limbus fly-through.](#)

**Movie 3.** [Large collector channel unit.](#)

**Movie 4.** [Automated surface reconstruction](#).

**Supplementary Data 1**. [Small caliber branches of the outflow vasculature.](#)